\documentclass[aps,prl,twocolumn,showpacs,superscriptaddress]{revtex4}

\usepackage{graphicx} % Include figure files

\begin{document}

\title
{Two- and one-dimensional honeycomb structures of silicon and germanium}

\author{S. Cahangirov}
\affiliation{UNAM-Institute of Materials Science and
Nanotechnology, Bilkent University, Ankara 06800, Turkey}
\author{M. Topsakal}
\affiliation{UNAM-Institute of Materials Science and
Nanotechnology, Bilkent University, Ankara 06800, Turkey}
\author{E. Akt{\"u}rk}
\affiliation{UNAM-Institute of Materials Science and
Nanotechnology, Bilkent University, Ankara 06800, Turkey}
\author{H. \c{S}ahin}
\affiliation{UNAM-Institute of Materials Science and
Nanotechnology, Bilkent University, Ankara 06800, Turkey}
\author{S. Ciraci} \email{ciraci@fen.bilkent.edu.tr}
\affiliation{UNAM-Institute of Materials Science
and Nanotechnology, Bilkent University, Ankara 06800, Turkey}
\affiliation{Department of Physics, Bilkent University, Ankara
06800, Turkey}

\date{\today}

\begin{abstract}
First-principles calculations of structure optimization,
phonon modes and finite temperature molecular dynamics predict
that silicon and germanium can have stable, two-dimensional, low
-buckled, honeycomb structures. Similar to graphene, these puckered structures are ambipolar and their charge carriers can behave like a massless Dirac fermions due to their $\pi$- and $\pi^{*}$-bands which are
crossed linearly at the Fermi level. In addition to these
fundamental properties, bare and hydrogen passivated nanoribbons
of Si and Ge show remarkable electronic and magnetic properties,
which are size and orientation dependent. These properties offer
interesting alternatives for the engineering of diverse
nanodevices.
\end{abstract}

\pacs{73.22.-f, 63.22.-m, 61.48.De} \maketitle

The unusual electronic properties of graphene, which is derived
from its planar honeycomb structure leads to charge carriers
resembling massless Dirac fermions \cite{wallace}. Recent
synthesis of graphene \cite{novo,zhang,berger} has demonstrated
that this truly two dimensional (2D) structure is stable and has
introduced novel concepts \cite{novo,kats,geim}. Not only the
fundamental properties of 2D graphene, but also interesting size
and geometry dependent electronic and magnetic properties of their
quasi 1D nanoribbons \cite{ribbon,dai} have been revealed. While
the research interest in graphene and its ribbons is growing
rapidly, one has started to question whether the other Group IV
elements in Periodic Table, such as Si and Ge, have stable
honeycomb structure \cite{takeda,durgun}. Even before the
synthesis of isolated graphene, ab-initio studies based on the
minimization of the total energy has revealed that a buckled
honeycomb structure of Si can exists \cite{takeda,durgun}.

In this letter, based on the state-of-the-art structure
optimization, phonon dispersion and ab-initio finite temperature
molecular dynamics calculations within density functional theory (DFT) 
\cite{method} we show that the low-buckled honeycomb structures 
of Si and Ge can be stable. Their band structures show linear band crossing at the Fermi level and thus have Dirac points at the K- and K$^{\prime}$-points of the hexagonal Brillouin zone (BZ), even for puckered structure. These results are somehow unexpected but may have important consequences. For example, the bands display an ambipolar character and the charge carriers behave like massless Dirac fermions in a small energy range around the Fermi level, $E_F$. Even more remarkable is that the armchair and zigzag nanoribbons of Si and Ge can exist and display unusual properties which are crucial for future device applications.

\begin{figure}
\includegraphics[width=8.4cm]{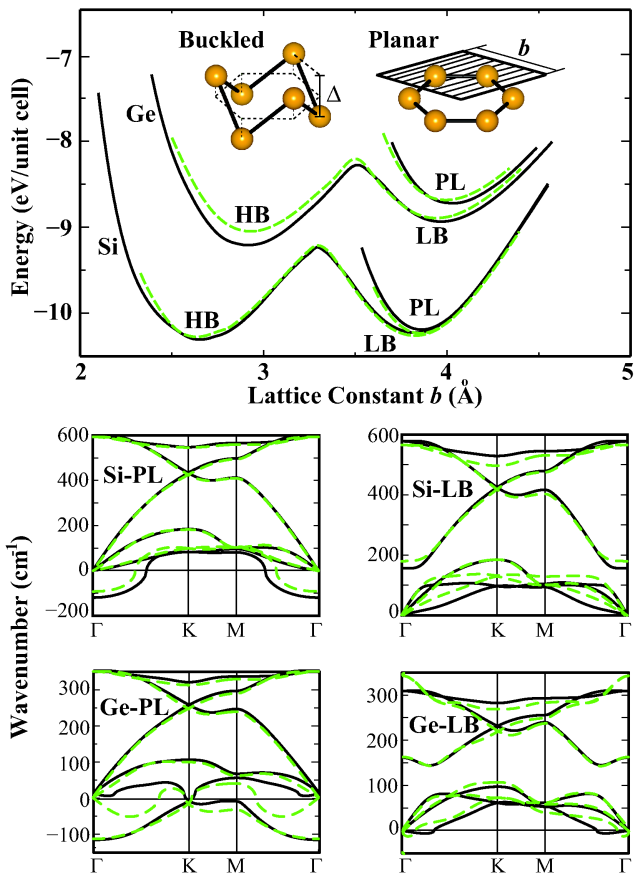}
\caption{(Color online) Upper panel: Energy versus hexagonal
lattice constant of 2D Si and Ge are calculated for various
honeycomb structures. Black (dark) and dashed green (dashed light)
curves of energy are calculated by LDA using PAW potential and
ultrasoft pseudopotentials, respectively. Planar and buckled
geometries together with buckling distance $\Delta$ and lattice
constant of the hexagonal primitive unit cell, $b$ are shown by
inset. Lower panels: Phonon dispersion curves obtained by
force-constant and linear response theory are presented by black
(dark) and dashed green (dashed light) curves, respectively.}
\label{fig:figure1}
\end{figure}

Calculated variation of the binding energy of the relaxed
honeycomb structure of Si and Ge as a function of the lattice
constant is presented in Fig.\ref{fig:figure1}. Here planar (PL),
low-buckled (LB) and high buckled (HB) honeycomb structures
correspond to distinct minima. The PL honeycomb structure is the
least energetic configuration and is not stable. The important
question to be addressed is whether these puckered LB and HB geometries
correspond to real local minima in the Born-Oppenheimer surface. 

PL structure of Si and Ge have phonon modes, which have imaginary frequencies in BZ. For the minimum energy PL structure of Si,
optical and acoustical branches hybridize and one optical (ZO) branch
is lowered into acoustical frequencies and have imaginary
frequencies along $\Gamma-K$ direction of BZ. The situation for PL
Ge structure is even more dramatic with one acoustical and one
optical branch having imaginary frequencies. As for HB honeycomb
structures of Si and Ge with a buckling of $\Delta_{HB}$~$\approx$~2~\AA, they have also imaginary phonon frequencies for a large portion of BZ. Moreover, structure optimization of HB structure on the (2$\times$2) supercell results in an instability with a tendency towards clustering. Clearly, the unstable HB structure does not correspond to a real local minimum; it can occur only under the constraint of the (1$\times$1)
hexagonal lattice.

The phonon dispersion curves in Fig.\ref{fig:figure1} indicate
that 2D periodic LB honeycomb structure of Si is stable. With an
equilibrium buckling $\Delta_{LB}$=0.44 \AA, its optical and
acoustical branches are well separated and all branches have
positive frequency. Two acoustical branches are linear as
\textbf{k} $\rightarrow$ 0. Whereas the transverse branch
displays a quadratic dispersion near $\Gamma$-point, since the
force constants related with the transverse motion of atoms
decays rapidly \cite{liu}. The phonon dispersion curves of 2D
periodic LB structure of Ge having a buckling of
$\Delta_{LB}$=0.64 \AA~are similar to those of Si, except the
frequencies of Ge are almost halved due to relatively smaller
force constants. The transverse acoustical phonon branch has
imaginary frequencies near $\Gamma$-point. This may be interpreted
as LB structure of Ge can be unstable if the wavelength of this
particular mode $\lambda > 3b$ ($7b$ according to phonon
dispersion curves calculated using DFPT \cite{method}), whereas
its flakes can be stable. In fact, the structure optimization on
the $(l \times l)$ supercells (where for $l=2-8$ atoms are
displaced along random directions from their optimized positions
and subsequently the structure is relaxed) is resulted in the
atomic configuration with periodic rippling for $l>3$. The
stability of LB structures of Si and Ge are further tested by
extensive ab-initio finite temperature molecular dynamics
calculations using time steps of $\delta t= 2 \times 10^{-15}$
seconds. In these calculations the $(4\times4)$ supercell is used
to lift the constraint of $(1\times1)$ cell. Periodic 2D LB 
structure of Si (Ge) is not destroyed by raising the temperature from T=0 K to 1000 K (800 K) in 100 steps, and holding it at T=1000 K (800K) for 10 
picoseconds (ps). A finite size, large hexagonal LB flake of Si (Ge) with
hydrogen passivated edge atoms is not destroyed upon raising its
temperature from 0 K to 1000 K (800 K) in 100 steps and holding it for more than 3 ps.

\begin{table}
\caption{Binding energy and structural parameters calculated for
the bulk and 2D Si and Ge crystals. $a_{\it{bulk}}$ [in \AA],
$E_{\it{c,bulk}}$ [in eV per atom], $\it{\Delta_{HB}}$ [in \AA],
$\it{\Delta_{LB}}$, $b_{LB}$, $d_{LB}$ and $E_{c,LB}$, respectively, stand for bulk cubic lattice constant, bulk cohesive energy, high-buckling distance, low-buckling distance, hexagonal lattice constant of 2D LB honeycomb structure, corresponding nearest neighbor distance and corresponding cohesive
energy.} \label{tab:bare}
\begin{center}
\begin{tabular}{cccccccc}
\hline  \hline
&a$_{bulk}$\space\space\space &$E_{c,bulk}$\space\space\space & $\triangle_{HB}$\space\space\space & $\triangle_{LB}$\space\space\space & $b_{LB}$\space\space\space & $d_{LB}$\space\space\space & $E_{c,LB}$\space\space\space \\
\hline
\textbf{Si}   & 5.41 &  5.92 & 2.13  &  0.44  & 3.83   & 2.25   &    5.16     \\
\hline
\textbf{Ge}   & 5.64 &  5.14 & 2.23  &  0.64  & 3.97    & 2.38   &   4.15      \\
\hline
\end{tabular}
\end{center}
\end{table}

We believe that the present analysis together with calculated
phonon dispersion curves provides stringent test
for the stability of LB honeycomb structure of both Si and Ge. In
this respect, LB structures of Si and Ge appear to be a contrast
to 2D C and BN forming only stable planar honeycomb structure
\cite{sahin}. The situation with three different minima
corresponding to PL, LB and HB geometries of 2D Si and Ge in
Fig.\ref{fig:figure1} is reminiscent of those of 1D atomic chains.
Earlier, it has been shown that while several elements and III-V
compounds form linear, wide-angle (i.e. LB) and low-angle (i.e.
HB) atomic chains \cite{sen,senger}, only C and BN form stable
linear atomic chains \cite{senger,tongay}. That C and BN form
linear 1D atomic chains and 2D planar honeycomb structures arises
from the strong $\pi$-bonding. Despite the weakened $\pi$-bonding, the stability of Si and Ge LB structures are maintained by puckering induced dehybridization. As a result, the perpendicular $p_z$-orbital, which forms $\pi$-bonding and hence $\pi$- and $\pi^{*}$-bands, combines with the $s$-orbital. Relevant lattice parameters and cohesive energies of LB Si and Ge honeycomb structures are given in Table \ref{tab:bare}. Different potentials (PAW or ultrasoft pseudopotential) \cite{method} yielded values which differ only 1\%.

\begin{figure}
\includegraphics[width=8.4cm]{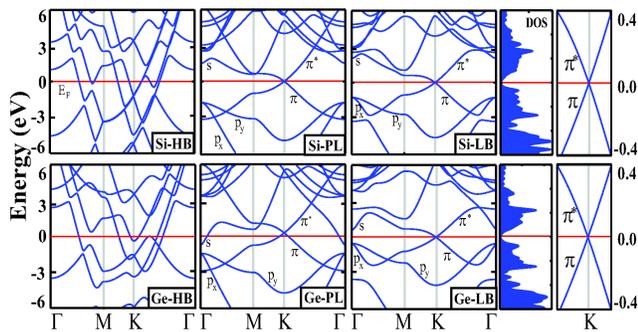}
\caption{(Color online) Energy band structure of Si and Ge are
calculated for high-buckled (HB), planar (PL) and low-buckled (LB)
structures. For LB structure the density of states (DOS) is also
presented. The crossing of the $\pi$- and $\pi^{*}$- bands at K- and
K$^{\prime}$-points of BZ is amplified to show that they are
linear near the cross section point.  Zero of energy is set at the
Fermi level, $E_F$. $s$, $p_{x,y}$ orbital character of bands are
indicated.} \label{fig:figure2}
\end{figure}

\begin{figure}
\includegraphics[width=8.4cm]{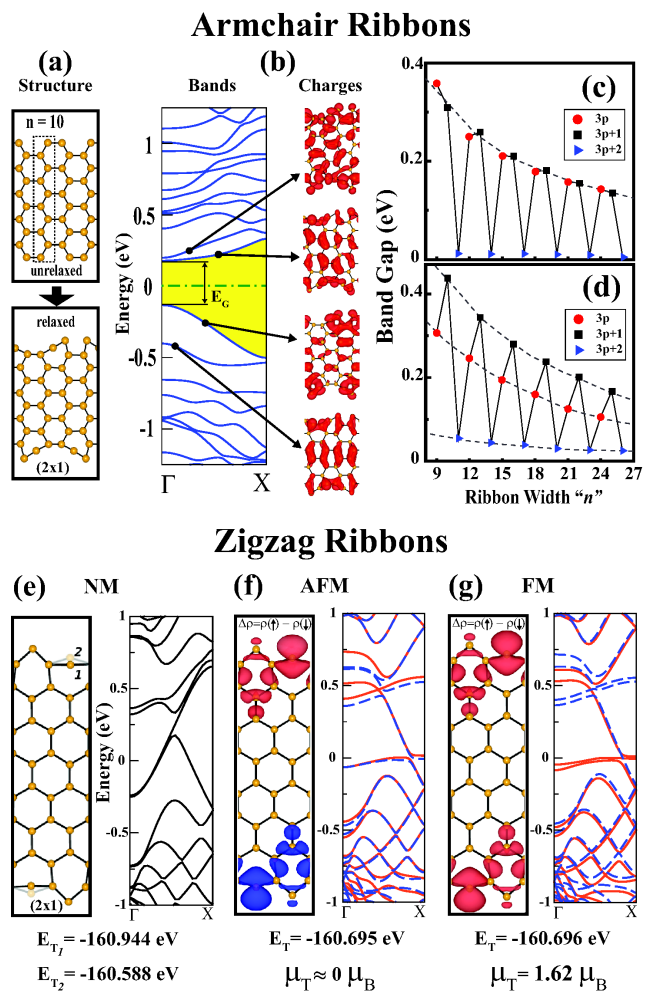}
\caption{(Color online) Ideal and relaxed atomic structure
displaying a ($2\times1$) asymmetric dimer like reconstruction (a);
electronic energy bands and isosurface charge density of selected
states (b); variation of band gap, $E_G$ with the width $n$ of
bare Si armchair nanoribbons (c); and similar variation of $E_G$
for the hydrogen saturated nanoribbons showing oscillations
depending on whether $n=3p$, $3p+1$ or $3p+2$ ($p$ being an
integer) (d). Bare Si zigzag nanoribbons showing two different
($2\times1$) reconstruction geometries \cite{haneman} indicated by
"1" and "2" and the band structure of metallic non-magnetic (NM)
ground state corresponding to "1" (e). Isosurfaces of spin
density difference [$\Delta \rho = \rho^{\uparrow} -
\rho^{\downarrow}$] for spin-up (red/light) and spin-down
(blue/dark) states in different magnetic excited states together
with spin-up (solid-red/light) and spin-down (dashed-blue/dark)
bands: Antiferromagnetic (AFM) state (f); and ferromagnetic (FM)
state (g) together with their calculated total energies and
magnetic moments. Zero of the energy is set to $E_F$.}
\label{fig:figure3}
\end{figure}

The calculated electronic band structures and corresponding
density of states (DOS) of LB Si and Ge are presented in
Fig.\ref{fig:figure2}. For the sake of comparison bands for
unstable planar and HB structures are also given. Two dimensional
HB Si and Ge are metallic. The bands of PL and LB structures are
similar except that specific degeneracies split due to lowering of
point group rotation symmetry from C$_6$ in PL geometry to C$_3$
in LB geometry. Similar to graphene, $\pi$- and $\pi^{*}$-bands of
LB Si crossing at K- and $K^{'}$-points at $E_{F}$ are
semimetallic. For PL Ge, since the $s$-like lowest conduction band
of planar Ge dips into the Fermi level, $\pi$- and $\pi^{*}$-bands
cross at K- and K$^{\prime}$-points above the Fermi level.
Therefore, PL structure of Ge  is metallic. Upon a structural
transformation from PL to LB structure, the crossing point of the
$\pi$- and $\pi^{*}$-bands of Ge shifts to Fermi level. At the
end, LB structure of Ge becomes also semimetallic. Around the
crossing point, these bands are linear. This behavior of bands, in
turn, attributes a massless Dirac fermion character to the charge
carriers. Interestingly, by neglecting the second and higher order
terms with respect to $q^{2}$, the Fermi velocity is estimated to
be $v_{F} \sim 10^{6} m/s$ for both Si and Ge by fitting the
$\pi$- and $\pi^{*}$- bands at $\bf{k}=\bf{K}+\bf{q}$ to the
expression

\begin{equation}
v_{F} \simeq E(\bf{q})/\hbar|\bf{q}|
\end{equation}

We note that $v_{F}$ calculated for LB honeycomb structures of Si
and Ge are rather high and close to that calculated for graphene
using the tight-binding bands. In addition, because of the
electron-hole symmetry at K- and $K^{\prime}$-points of BZ, LB Si
and Ge are ambipolar for E{(\bf{q})}= $E_{F} \pm \epsilon$,
$\epsilon$ being small. The ambipolar effect and high $v_{F}$ of
LB Si and Ge are remarkable properties.

While LB crystals of Si and Ge are of fundamental importance, any
application involving these materials requires only a small piece
of them or a flake, but not an infinite size. In this respect,
their ribbons of nanometer scale with well-defined shape may be
crucial for device applications. Whether nanoribbons of Si and Ge
show behaviors similar to graphene is the next question to be
answered. Here we consider Si and Ge armchair and zigzag
nanoribbons of different widths, in terms of the number of Si or Ge
atoms $n$ forming a continuous chain between two edges. The
ribbons having width $n>7$, $\sim$1 nm, preserve their LB
honeycomb structure upon structure relaxation. The value of the
buckling decrease near the edges. Their both edges, undergo a
(2$\times$1) reconstruction, which is different for different
orientation. Whereas the reconstruction disappear when the
dangling bonds at the edges are terminated by hydrogen atoms.

In Fig.\ref{fig:figure3}(a), we show the minimum energy
reconstruction pattern of $n=10$ armchair nanoribbons among four
other ($2\times1$) patterns. Si armchair nanoribbons are
nonmagnetic semiconductors with band gaps relatively smaller than
those of graphene. Generally, owing to quantum confinement effect
the band gap $E_G$ increases with decreasing width $n$. However,
similar to graphene, the variation of $E_G$ with $n$ shows an
oscillatory (or family) behavior. For example, if $n=3p+2$ ($p$
being an integer), $E_G$ is very small, but it is large for $n=3p$
and $n=3p+1$. Upon the saturation of dangling bonds with hydrogen,
the value of the band gap increases for small $n$, but continues
to show the "oscillatory behavior". In this case also $E_G$ is
still small for $n=3p+2$. Similar oscillatory behavior is also
calculated for Ge armchair nanoribbons. We note that the DFT may underestimate the calculated band gaps. The variation of the band gap with $n$ is an important property, which may lead to formation of quantum dot or multiple quantum wells through the width modulation \cite{admin}.

We performed spin-dependent total energy and electronic structure
calculations for bare and hydrogen terminated zigzag nanoribbons.
In the (2$\times$1) reconstruction of bare zigzag nanoribbons in
Fig.\ref{fig:figure3}(e), one Si atom at the edge is pushed down
while the adjacent atom is raised. This situation is reminiscent
of the (2$\times$1) reconstruction of Si(111) surface pointed out
earlier by Haneman \cite{haneman}. In Fig.\ref{fig:figure3}(e) one
distinguishes however, out-of-plane and in-plane reconstruction
geometries indicated by "1" and "2", respectively. We found that,
the out-of-plane ($2\times1$) reconstruction geometry has a
nonmagnetic (NM), metallic ground state. However, metallic
antiferromagnetic (AFM) and ferromagnetic (FM) states in
Fig.\ref{fig:figure3}(f) and \ref{fig:figure3}(g), respectively,
are excited states. This situation is, however, reversed upon the
termination of dangling bonds by hydrogen; namely magnetic states
have lower energies than NM state. The analysis of the difference
charge density, $\Delta \rho = \rho^{\uparrow} -
\rho^{\downarrow}$ in Fig.\ref{fig:figure3}(f) and
\ref{fig:figure3}(g) indicates that the AFM and FM states have
almost equal energies, the FM state being 1 meV more energetic.
This energy difference between AFM and FM states, which is within
the accuracy limits of DFT and hence is not decisive, is however
reversed, if noncollinear calculations including spin-orbit
interaction are performed. In the AFM case, the edge states have
opposite spins and approximately zero magnetic moment in the unit
cell. On the other hand, the FM state has magnetic moment of 1.62
$\mu_B$. In both AFM and FM state in Fig.\ref{fig:figure3}, the
lowered edge Si atoms have larger magnetic moment, since each
raised edge atom donates electrons to adjacent lowered edge atoms.

In conclusion, calculations based on DFT show that Si and Ge can remain stable in LB honeycomb structure, which attribute them important properties similar to graphene. Armchair and zigzag nanoribbons of LB Si and Ge in honeycomb structure exhibit electronic and magnetic
properties, which depend strongly on their size and geometry. The
electronic properties of these nanoribbons undergo dramatic change
depending whether their edges are passivated by hydrogen. These
properties of Si and Ge nanoribbons can be used for diverse
device applications. Further to the predictions of the present
study, recent work by Nakano \cite{nakano} \textit{et al.} achieving the
soft synthesis of single Si monolayer sheet on a substrate holds
the promise for the synthesis of Si and Ge nanoribbons having honeycomb
structures.

We acknowledge stimulating discussions with Can Ataca.
Part of the computations have been carried out by using UYBHM at
Istanbul Technical University through a grant (2-024-2007).


\begin{thebibliography}{99}
\bibitem{sahin}
We note the tendency that BN, SiC, AlN, GaN form planar honeycomb
structure, while some of III-V and II-VI compounds are buckled.
\end{thebibliography}
\end{document}